# Robust Speech and Natural Language Processing Models for Depression Screening

*Y. Lu, A. Harati, T. Rutowski, R. Oliveira, P. Chlebek, E. Shriberg*

Ellipsis Health, San Francisco, California, USA
{yang,amir,tomek,ricardo,piotr,liz}@ellipsishealth.com

**Abstract** — Depression is a global health concern with a critical need for increased patient screening. Speech technology offers advantages for remote screening but must perform robustly across patients. We have described two deep learning models developed for this purpose. One model is based on acoustics; the other is based on natural language processing. Both models employ transfer learning. Data from a depression-labeled corpus in which 11,000 unique users interacted with a human-machine application using conversational speech is used. Results on binary depression classification have shown that both models perform at or above AUC=0.80 on unseen data with no speaker overlap. Performance is further analyzed as a function of test subset characteristics, finding that the models are generally robust over speaker and session variables. We conclude that models based on these approaches offer promise for generalized automated depression screening.

## I. Introduction

Depression is a debilitating condition globally [1][2]. A key challenge in its management is screening so that patients at risk can be identified early for treatment. Spoken language offers promise for automated depression screening because it is used in human-human conversations with clinicians, is engaging for users, and can be cheaply and conveniently collected remotely using personal devices.

Prior research shows that speech contains cues to depression risk, both in what is said (word patterns) and how it is said (acoustic-prosodic patterns) [3][4]. There is a growing interest in developing speech technology in this domain [5][6][7]; community participation has increased with shared data sets [8]−[11]. Despite these advances, there is little work on large corpora because of challenges in both collecting and sharing behavioral health data.

Because speech and language are themselves behavioral and speaker-dependent, it is critical to understand how real-world variables associated with the patient or context affect system performance. Past work on smaller data sets has not allowed for the study of these questions. To this end, we use a corpus of over 11,000 unique speakers to examine the effects of basic metadata on the performance of both an acoustic model and a word-based model for depression risk prediction. Both models use transfer learning to augment model training. Each model offers different practicality, portability, and privacy advantages. While models can be fused to optimize overall performance, we present them individually to directly assess differences in robustness patterns.

We report the performance of each model overall as well as on subsets of test data *without model retraining or optimization for each subset*. While machine learning offers powerful ways to retrain or adapt models, fielding simpler systems requiring less labeled training data is currently more practical. Our goal is thus to assess how well deep learning models trained on a fixed set of samples will generalize over specific speaker and temporal variables in unseen data.

## II. Data

### A. Speech corpus

We use a depression-labeled corpus of American English speech collected by Ellipsis Heath. Speakers range from 18–65 years old with a mean age of roughly 30 years. Users interacted with an application that prompted them to speak about their personal lives regarding topics such as self-care, relationships, and home life. The total session lengths averaged 4–5 minutes. In each session, users also completed a Patient Health Questionnaire-8 (PHQ-8) [12], an instrument assessing self-reported depression severity. The distribution of PHQ-8 values in our corpus is shown in Figure 1.

Following [12], PHQ-8 scores of 10 and above were mapped to positive for depression (Dep+) and those below 10 were mapped to negative (Dep-). We used these binary class labels as a gold standard to train and evaluate model performance.

The corpus contains roughly 16,000 sessions from roughly 11,000 unique speakers. Data was partitioned into train and test sets with no overlapping speakers.

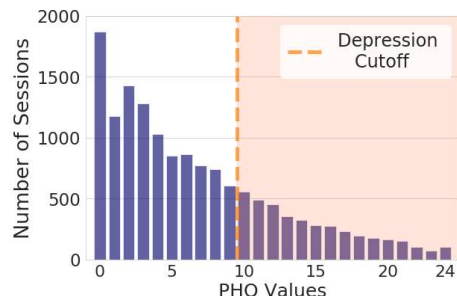

Figure 1. Distribution of PHQ-8 labels for Ellipsis corpus.

Table 1. Corpus statistics by partition and depression class

|  | Total | Train Dep- | Train Dep+ | Test Dep- | Test Dep+ |
|---|---|---|---|---|---|
| **Sessions** | 15950 | 9266 | 3606 | 2425 | 653 |
| **Hours** | 1130 | 795 | 335 | 234 | 69 |
| **Words** | 11.68M | 6.40M | 2.64M | 1.87M | 0.53M |

Speakers with more than one session were only used in the training set. Corpus statistics with partition information are provided in Table 1.

*B. User and session metadata*

The corpus contains session metadata including self-reported information about the user and automatically-generated information about session timing. We focus here on the following metadata: gender, age group, smoking habits, ethnicity, marital status, and location. We find two types of differences in priors over these characteristics. Firstly, priors differ in terms of their *frequency in the corpus*. For example, our corpus contains slightly more female-user sessions than male-user sessions. Secondly, priors differ in terms of their *distribution of gold standard labels (PHQ-8 values)*. For example, depression values for females are slightly higher than those for males. Model robustness requires generalizing over both types of differences in priors.

*Session metadata* includes information about the administration of the session. Here we include the following characteristics: local time of the day, day of the week, and season during which the session was recorded. In our collection, we allowed users to choose their recording days and times. We found differences in both the frequency of recording and the PHQ-8 value priors. As an example of the latter, we discovered that even after averaging over more than 16,000 sessions and 11,000 users, sessions that occurred at certain times of day (local to the user's time zone) were more likely to be depressed than sessions collected at other times of day. This pattern varies relatively smoothly, as shown in Figure 2.

### III. DEEP LEARNING MODELS

To predict depression class, two deep learning models were used, both of which incorporate transfer learning. The models are trained as binary classifiers based on the class mapping shown in Figure 1. To examine the effects of metadata on each model type, results are included for each model individually rather than fused. Due to the large number of results reported, we use only single-model versions of both the acoustic and natural language processing (NLP) systems. We note that for both models, performance improvements are achieved using within-model-type fusion of multiple systems.

*A. Acoustic model*

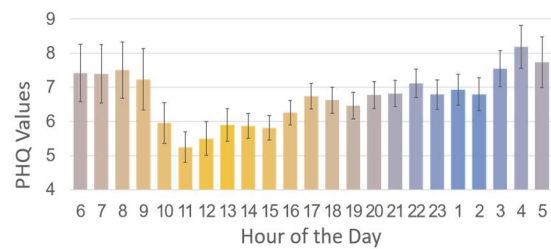

Figure 2. PHQ-8 value (mean and variance) for sessions by time of day the session was recorded (local time), for full corpus.

To capture how depression is reflected in acoustic-prosodic characteristics of the signal, we have designed an acoustic model. This model consists of layers of convolutional neural networks (CNNs) [13] and long short-term memories (LSTMs) [14] (collectively the "encoder"), and a prediction layer. As shown in Figure 3, the model can be broken down into three parts: an acoustic feature extractor, an encoder, and a predictor.

The inputs of the model are filter-bank coefficients. These have been computed at a 10-millisecond frame rate over a window of 25 milliseconds for fixed-length segments of 25 seconds (optimized empirically). The encoder has been used to project the input features into a representation that is used by the predictor layer. We have used an automatic speech recognition (ASR) task for transfer learning. During transfer learning, a decoder network is attached to the encoder to train the encoder and decoder using ASR data. Once the encoder has been trained, we remove the decoder and train the prediction layer using data with depression labels while fixing the weights of the encoder. The predictions for all speech regions belonging to one session are then gathered and fed into another neural network to generate one depression prediction over the full session.

*B. NLP model*

To capture word patterns associated with depression, we used an NLP deep learning model. We have found good results given our resource constraints using an average stochastic gradient descent (ASGD) weight dropped

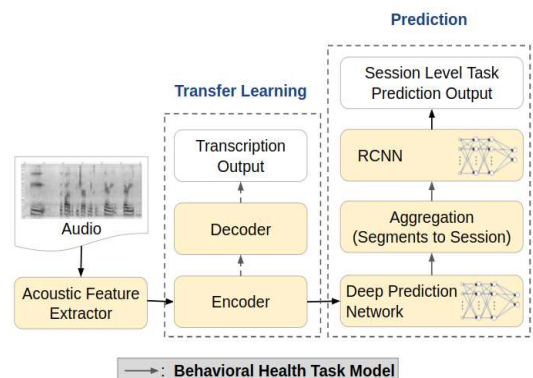

Figure 3. Overview of acoustic model.

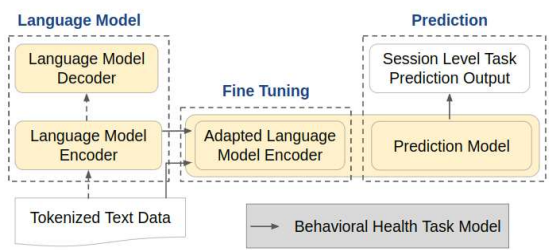

Figure 4. Overview of NLP model.

LSTM (AWD-LSTM) language model [15]. We follow the work described in [16] on the ULMFiT method. This approach uses the entire language model in transfer learning [17], rather than only the selected layers of the network. An overview is shown in Figure 4.

The generic language model is first trained using large amounts of text data from publicly-available corpora commonly used for this type of pre-training, including Wikipedia [18]. Text data for this step as well as for future steps is tokenized using the Spacy tokenizer [19]. The generic language model is then further pre-trained on data closer in style to our behavioral health task domain. For this purpose, we used text data collected from sources such as health forums. In separate experiments, we find that including this step helps to stabilize our system's final behavioral health task predictions.

We then perform an adaptation step in which proprietary data without labels is used to further train the adapted language model encoder. Using our behavioral health session and labels, additional layers are then added to perform either classification or regression, resulting in the prediction model. We note that at least two methods in [16] have helped us achieve good prediction results; these include gradually unfreezing layers and allowing different learning rate ratios for different layers.

IV. RESULTS AND DISCUSSION

We first analyze the performance of the acoustic and NLP models on all sessions in the test set. Since our task is binary classification, we show the trade-offs of sensitivity and specificity. Due to the number of fusion models, we show the results of a single NLP model and a single Acoustic model in Figure 5. For reference, data from three separate studies are also indicated. These data reflect results on depression detection performance by Primary Care Providers (PCPs) not specifically trained in mental health assessment. Note that these studies are not directly comparable to ours, nor to each other, due to differences in the data and methods used. They do, however, provide a crude indication of human performance in the context of a general PCP visit [20][21][22].

As shown in Figure 5, both the single acoustic model and NLP model achieve an area under the curve (AUC) of close to or above 0.80. Within-model fusion gives us an additional 2-3% in AUC performance. These systems use no information other than the speech sample itself; that is, no metadata, patient history, or other information (such as visual information) is used for the acoustic and NLP results. The NLP model performs better overall than the acoustic system, but both systems show strong results in line with or better than the PCP reference studies.

Table 2 shows results using our acoustic and NLP models on subsets of the test data broken down by metadata categories. Session counts and depression rates are also indicated for each entry. We excluded categories with session counts below 150 to avoid noisy results.

To compare AUC values and evaluate significant differences, we employ the DeLong test [23], a non-parametric test for comparing an AUC of two or more correlated receiver operation characteristic (ROC) curves. This test is appropriate for our setup; we note that we have no nesting of model inputs in our framework. Due to the relatively large samples sizes in some data categories, we used the fast implementation by [24] to reduce computation time.

The following observations can be made from Table 2. Overall, both the acoustic and NLP models are robust over speaker characteristics. There are some exceptions, as indicated by entries marked with an asterisk. A marked entry indicates that the AUC of the category is significantly higher or lower than that of other members in the metadata group with $p<0.05$.

We see that for the acoustic model, performance on depression classification is significantly lower if the speaker is aged 26–35 or identifies as Hispanic. There are no such differences for the NLP model. We also collected metadata on user location because it is correlated with other factors such as socioeconomic status, regional accents (relevant to ASR), and other variables. The rate of depression is similar across locations, but there are performance differences. It is premature for us to propose reasons for this but we include this set to demonstrate that not all variables result in similar performance.

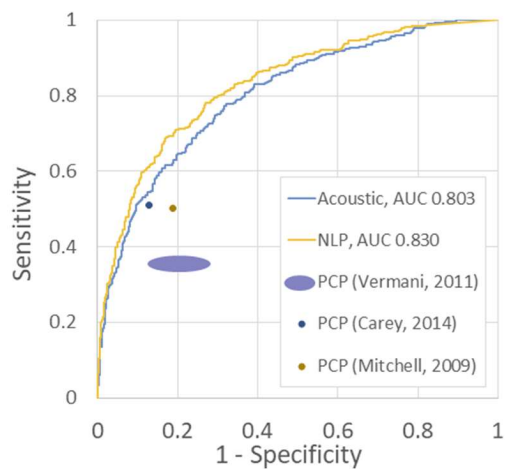

Figure 5. ROC curves for a single Ellipsis acoustic model and a single Ellipsis NLP model. Circle markers denote the point of equal error rate for sensitivity and specificity on each curve. For reference, results from human PCP studies ([20][21][22] respectively) are shown but are not directly comparable due to differences in data and methods used.

Table 2. Model Performance by Test Subset.

| | Metadata | Categories | Train set Sess. Count | Test set Sess. Count | Depression Rate | Mean PHQ | Acous. AUC | NLP AUC |
|---|---|---|---|---|---|---|---|---|
| **Base performance over all test set** | | | 11 215 | 3080 | 25.7% | 5.93 | 0.779 | 0.825 |
| User Metadata | Gender | Male | 3125 | 1244 | 20.4% | 5.74 | 0.769 | 0.819 |
| | | Female | 4419 | 1790 | 35.3% | 6.77 | 0.774 | 0.820 |
| | Age group | 18-25 | 2087 | 847 | 30.0% | 7.32 | 0.792 | 0.828 |
| | | 26-35 | 3256 | 1382 | 24.8% | 6.40 | 0.752* | 0.820 |
| | | 36-45 | 1444 | 513 | 18.7% | 5.60 | 0.790 | 0.808 |
| | | 46-65 | 766 | 283 | 34.6% | 4.78 | 0.792 | 0.819 |
| | Smoking | Non-smoker | 3850 | 813 | 23.2% | 6.44 | 0.803 | 0.836 |
| | | Smoker | 1807 | 397 | 31.3% | 7.47 | 0.767 | 0.808 |
| | US States (selected) | California | 924 | 266 | 26.8% | 6.68 | 0.741 | 0.830 |
| | | Florida | 831 | 253 | 26.2% | 6.41 | 0.842* | 0.875* |
| | | Texas | 723 | 232 | 26.0% | 6.66 | 0.810 | 0.845 |
| | | New York | 596 | 142 | 25.7% | 6.70 | 0.815 | 0.887* |
| | Ethnicity | Caucasian | 5219 | 2039 | 24.7% | 6.05 | 0.796 | 0.826 |
| | | African American | 569 | 241 | 19.7% | 5.63 | 0.777 | 0.812 |
| | | Hispanic | 552 | 248 | 25.0% | 6.73 | 0.676* | 0.788 |
| | | Asian American | 452 | 185 | 20.0% | 5.61 | 0.789 | 0.841 |
| | | Mixed | 364 | 173 | 31.3% | 7.22 | 0.768 | 0.827 |
| | Marital | Never married | 1850 | 188 | 31.5% | 7.84 | 0.778 | 0.857 |
| | | Married | 1220 | 173 | 21.2% | 5.20 | 0.774 | 0.829 |
| Session Metadata | Time of day[2] | Morning | 1275 | 476 | 24.3% | 6.39 | 0.785 | 0.823 |
| | | Afternoon | 3471 | 1127 | 22.7% | 6.08 | 0.776 | 0.841 |
| | | Night | 4012 | 1005 | 26.3% | 6.76 | 0.783 | 0.815 |
| | | Late night | 2457 | 472 | 31.2% | 7.26 | 0.758 | 0.804 |
| | Day of week | Weekdays | 9343 | 2307 | 25.2% | 6.54 | 0.782 | 0.832 |
| | | Weekends | 1872 | 773 | 27.5% | 6.88 | 0.772 | 0.802 |
| | Time of year[3] | Summer | 993 | 756 | 20.1% | 5.39 | 0.818 | 0.838 |
| | | Rest of the year | 6056 | 2324 | 27.0% | 6.88 | 0.769 | 0.821 |

\* $p<0.05$ in statistical significance test of AUC [23]. The denoted category has a significantly different AUC than others in same set.
[1] The mean PHQ is calculated using only the first session for each speaker in the designated subset to avoid skewing of mean PHQ by speakers who participated multiple times. For most speakers, the first session was their only session.
[2] We partitioned the day into four 6-hour parts for analysis: morning is defined as 6am to noon, afternoon as noon to 6pm and night as 6pm to midnight and late night as midnight to 6am. Time is local for the user. Other session factors such as day of week and month of year also showed effects but are not shown due to space restrictions.
[3] Summer is defined as the time of year where the month is June, July, or August.

Overall, both models are remarkably consistent. They discriminate between positive and negative depression classes at a level that is not significantly different from the other members of the set and similar to the overall performance level of the full test set, as shown in Table 2. Despite differences in priors and in PHQ-8 distributions, the ability of the models to separate classes does not change significantly for most user conditions, including gender, smoking, and marital status. The AUC of the 26–35 year age group is slightly lower than that of the rest of the population but this difference is statistically significant. We are currently investigating this result; given that this group has a large amount of training data, data sparsity is not a concern.

As discussed earlier in this paper, the AUC is robust over session metadata for the time of day. For example, in Figure 2, we saw changes in PHQ-8 average over the course of the day, but the AUC for depression detection did not change significantly over this factor. Both models are largely robust to other time-related session variables, including time of day, day of week, and time of year. This performance consistency holds despite differences across categories in PHQ-8 priors.

## V. CONCLUSIONS

We studied the performance of deep learning models for depression classification from natural speech using acoustic and NLP models without additional training or modifications for particular subsets. Both acoustic and NLP models take advantage of transfer learning in different ways. The models perform well when compared to reference data from human PCP studies, although we note that the data and methods used differ across these studies. The NLP model performs better than the acoustic model across operating points.

We further analyzed the robustness of each model by isolating the test data subsets based on user or session metadata. The results showed that with only a few exceptions for the acoustic system, the AUC was stable for the subsets. This pattern was true despite differences in priors for both the category frequencies and PHQ-8 value distributions of the metadata categories.

Future work should examine additional types of metadata. An important source of variability is recording quality, which varies when using diverse patient devices and software. ASR quality is important for the NLP system and can be affected by audio issues and by accented and nonnative speech. Cross-corpus studies are an additional focus for further research. This paper is a start but is limited because we examine only subsets within a matched collection. It is critical to look at corpora with large demographic differences as well as with differences in how speech is collected by an application. Overall, we conclude that speech technology offers promise for the remote screening of depression and related behavioral health conditions.